\documentclass{sig-alternate-2013}

\newfont{\mycrnotice}{ptmr8t at 7pt}
\newfont{\myconfname}{ptmri8t at 7pt}
\let\crnotice\mycrnotice%
\let\confname\myconfname%

\permission{Permission to make digital or hard copies of all or part of this work for personal or classroom use is granted without fee provided that copies are not made or distributed for profit or commercial advantage and that copies bear this notice and the full citation on the first page. Copyrights for components of this work owned by others than the author(s) must be honored. Abstracting with credit is permitted. To copy otherwise, or republish, to post on servers or to redistribute to lists, requires prior specific permission and/or a fee. Request permissions from permissions@acm.org.}
\conferenceinfo{SIGMETRICS'14,}{June 16--20, 2014, Austin, Texas, USA. \\
{\mycrnotice{Copyright is held by the owner/author(s). Publication rights licensed to ACM.}}}
\copyrightetc{ACM \the\acmcopyr}
\crdata{978-1-4503-2789-3/14/06\ ...\$15.00.\\
http://dx.doi.org/10.1145/2591971.2591985}

\usepackage{url}
\usepackage{graphicx}
\usepackage{multirow}

\def\addauthorsection{}

\newcommand{\lsc}{LSC}
\newcommand{\incomp}{IN}
\newcommand{\out}{OUT}
\newcommand{\intend}{IN-TENDRILS}
\newcommand{\outtend}{OUT-TENDRILS}
\newcommand{\bridges}{BRIDGES}
\newcommand{\other}{OTHER}
\newcommand{\disc}{DISCONNECTED}

\makeatletter
\newcommand\footnoteref[1]{\protected@xdef\@thefnmark{\ref{#1}}\@footnotemark}
\makeatother

\begin{document}

\clubpenalty = 100000
\widowpenalty = 100000
\CopyrightYear{2014}
\title{Studying Social Networks at Scale:\\ Macroscopic Anatomy of the Twitter Social Graph}

 \numberofauthors{3} 
 \author{
 \alignauthor
 Maksym Gabielkov\\
        \affaddr{Inria}\\
       \affaddr{Sophia Antipolis, France}\\
 \email{maksym.gabielkov@inria.fr}
 \alignauthor
 Ashwin Rao\\
        \affaddr{Inria}\\
       \affaddr{Sophia Antipolis, France}\\
 \email{ashwin.rao@inria.fr}
 \alignauthor
 Arnaud Legout\\
        \affaddr{Inria}\\
       \affaddr{Sophia Antipolis, France}\\
 \email{arnaud.legout@inria.fr}
 }

\maketitle
\sloppy

\begin{abstract}
Twitter is one of the largest social networks using exclusively directed links among accounts.
This makes the Twitter social graph much closer to the social graph supporting real life communications than, for instance, Facebook.
Therefore, understanding the structure of the Twitter social graph is interesting not only for computer scientists, but also for researchers in other fields, such as sociologists.
However, little is known about how the information propagation in Twitter is constrained by its inner structure.

In this paper, we present an in-depth study of the macroscopic structure of the Twitter social graph unveiling the highways on which tweets propagate, the specific user activity associated with each component of this macroscopic structure, and the evolution of this macroscopic structure with time for the past 6 years.
For this study, we crawled Twitter to retrieve all accounts and all social relationships (follow links) among accounts; the crawl completed in July 2012 with 505 million accounts interconnected by 23 billion links. 
Then, we present a methodology to unveil the macroscopic structure of the Twitter social graph. This macroscopic structure consists of 8 components defined by their connectivity characteristics. Each component group users with a specific usage of Twitter. 
For instance, we identified components gathering together spammers, or celebrities.
Finally, we present a method to approximate the macroscopic structure of the Twitter social graph in the past, validate this method using old datasets, and discuss the evolution of the macroscopic structure of the Twitter social graph during the past 6 years.
\end{abstract}


\section{Introduction}
\label{sec:introduction}
Twitter is one of the largest social networks with more than 500 million registered accounts.
However, it differs from other large social networks, such as Facebook and Google+, because it uses exclusively arcs among accounts\footnote{Arcs---that are directed edges---represent the follow relationship in Twitter. If A follows B, A receives tweets from B, but B will not receive tweets from A, unless B follows A.}.
Therefore, the way information propagates on Twitter is close to how information propagates in real life.
Indeed, real life communications are characterized by a high asymmetry between information producers (such as media, celebrities, etc.) and content consumers.
Consequently, understanding how information propagates on Twitter has implications beyond computer science.

However, studying information propagation on a large social network is a complex task.
Indeed, information propagation is a combination of two phenomena.
First, the content of the messages sent on the social network will determine its chance to be relayed.
Second, the structure of the social graph will constrain the propagation of messages.
In this paper, we specifically focus on how the structure of the Twitter social graph constrains the propagation of information.
This problem is important because its answer will unveil the highways used by the flows of information.
To achieve this goal, we need to overcome two challenges.
First, we need an up-to-date and complete social graph.
The most recent publicly available Twitter datasets are from 2009 \cite{KwakEtAl2010, ChaEtAl2010}, at that time Twitter was 10 times smaller than in July 2012.
Moreover, these datasets are not exhaustive, thus some subtle properties may not be visible.
Second, we need a methodology revealing the underlying social relationships among users, a methodology that scales for hundreds of millions of accounts and tens of billions of arcs.
Standard aggregate graph metrics such as degree distribution are of no help because we need to identify the highways of the graph followed by messages. Therefore, we need a methodology to both reduce the social graph and keep its main structure.

In this paper, we overcome these challenges and make the following specific contributions. 
\begin{enumerate}
\item We collected the entire Twitter social graph, representing 505 million accounts connected with 23 billion arcs.
To the best of our knowledge, this is the largest \textit{complete} social graph ever collected.
\item We unveil a macroscopic structure in the Twitter social graph that preserves the highways of information propagation. Our method extends the one of Broder \textit{et al.}~\cite{Broder2000} and can be applied  to any kind of directed social graph.
\item We show that not only the macroscopic structure of the Twitter social graph constrains information propagation, but that each component of the macrostructure corresponds to group of users with a specific usage of Twitter. In particular, we show that regular, abandoned, and malicious accounts are not uniformly spread on the components of the macroscopic structure of the Twitter social graph. This result is important to understand how Twitter is used, where users with a specific usage are, and how to sample Twitter without a significant bias. 
\item We present a simple methodology to explore the evolution of the macroscopic structure of Twitter with time, we validate this methodology, and show that old datasets from 2009 do not represent the current structure of the Twitter social graph.
We explore this time evolution to understand the changes in the usage of Twitter since its creation.
\end{enumerate}

The remainder of this paper is structured as follows.
In Section~\ref{sec:dataset}, we present our methodology to crawl Twitter and discuss the dataset we collected.
We present and discuss, in Section~\ref{analysis_method},  the notion of macroscopic structure, then we describe a methodology to unveil this macroscopic structure.
We present the result of applying this methodology to our dataset in Section~\ref{twitter2012}.
In Section~\ref{evolution}, we propose a simple approach to estimate
the evolution of the macroscopic structure of the graph with time, validate this approach, and discuss the evolution of the Twitter social graph from 2007 to 2012.
Finally, we present the related work in Section~\ref{sec:related-work}, and conclude in Section~\ref{concl}.

\section{Measuring Twitter at Scale}
\label{sec:dataset}
In this section, we describe the methodology used to crawl the Twitter social graph, some high level characteristics of the dataset, the limitations of our crawl, and the ethical issues.

\subsection{Crawling Methodology}
\label{sec:methodology-crawl}
In order to collect our dataset, we used the Twitter REST API version 1.0~\cite{twitterapi} to crawl the information about user accounts and arcs between users.
The main challenge of the crawl is that API requests are rate-limited; an unauthenticated host could make at most 150 requests per hour with that API.
However this limit could be overcome by using a whitelisted machine.
Twitter used to whitelist the servers of research teams and data-intensive services upon request, this service has been discontinued since February 2011, but existing whitelisted machines could still be used.
We used four whitelisted machines to perform our crawl, two machines with a rate limit of 20,000 requests per hour and two with 100,000 requests per hour.

We also implemented and deployed a distributed crawler on 550 machines of PlanetLab~\cite{planetlab}, doubling the crawling rate compared to whitelisted machines only.

We crawled Twitter by user ID, such numeric IDs are assigned for new accounts sequentially, but with gaps \cite{KrishnamurthyEtAl2008}.
Therefore, we first determined using a random polling that the largest assigned ID is lower than 800~million, then we divided the range from 1 to 800~million into chunks of 10,000 IDs.
We selected an upper bound (800~million) much larger than the largest observed ID to be sure to do not miss any account.

We performed our crawl from March 20, 2012 to July 24, 2012.
We implemented a crawler that assigns chunks of 10,000 IDs to each crawling machine.
Then, for a given chunk, each crawling machine performs two steps.
First, the machine makes 100 requests for 100 IDs, the maximum number of IDs the lookup method of the API accepts, using an API call \cite{russell201121}.
When an ID corresponds to a valid account, we retrieve public numerical, boolean and date information\footnote{The public information returned by the API call we make is described in this URL \url{https://dev.twitter.com/docs/platform-objects/users}. We note that the history of the published tweets is not part of it.}.
Second, the machine collects the list of followings for all non-protected and valid accounts with at least one following.

We now define the notions of following, followers, and protected accounts that we use in this paper.
Each Twitter account can have \emph{followings} and \emph{followers}.
An account receives all published tweets from its \emph{followings}, and all its \emph{followers} receive its tweets.
Tweets, and list of followers and followings, are by default visible to everyone.
However, users can make their account \emph{protected} which makes this information visible only to its followers. Furthermore, following a \emph{protected} account requires manual approval from its owner~\cite{protected}.

\subsection{Limitations of the crawl}
\label{sec:limitations-crawl}

\begin{figure}[t]
\centering
\includegraphics[scale=0.47]{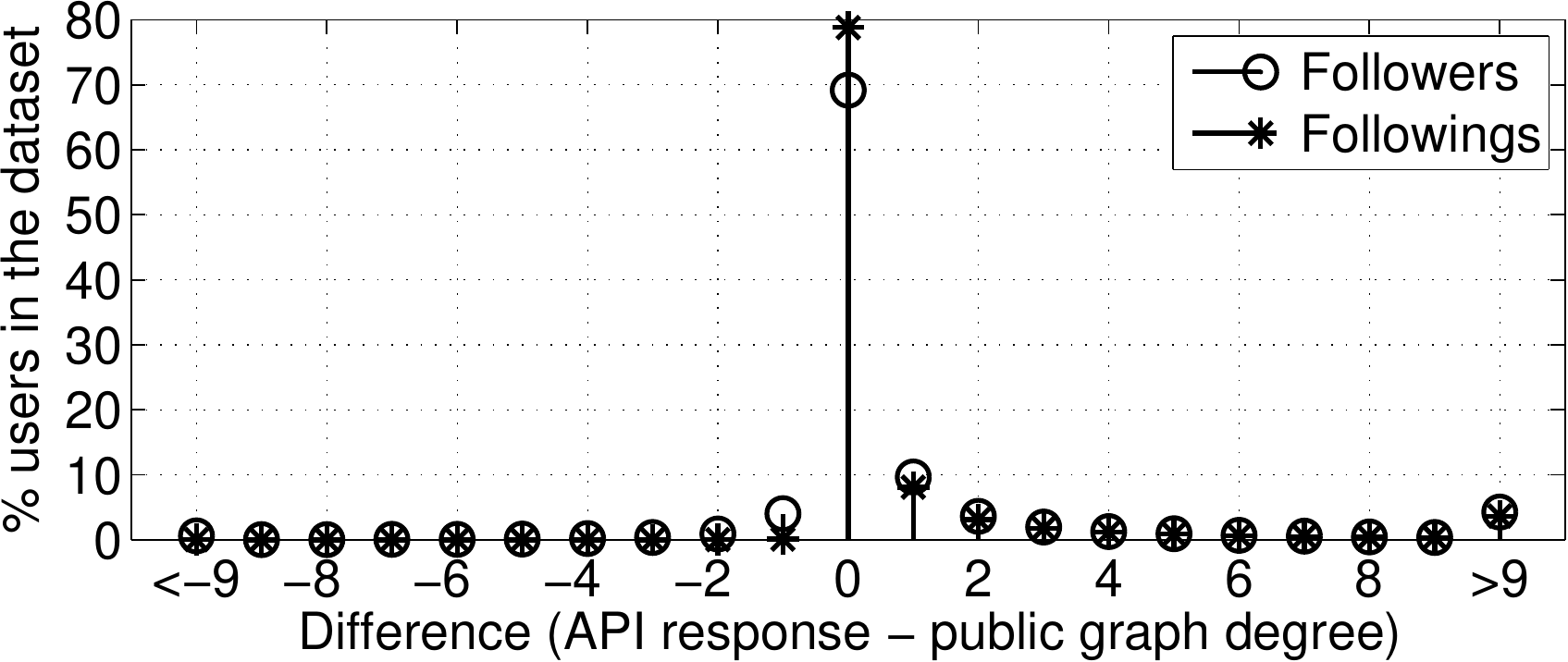}
\caption{
\textbf{The difference in number of followers and followings between the data from user accounts and the public social graph reconstructed from our dataset.}
}
\label{fig:diff}
\end{figure}

There are some accounts that we could not crawl, representing 6.33\% of the entire Twitter social graph.
We explain in the following the reasons why some accounts are not present in our dataset.

\begin{enumerate}
\item 32,112,668 accounts (5.97\% of the accounts in our dataset) are protected, so we cannot get their list of followings.
The degrees of nodes in the graph we analyzed do not take into account arcs to and from protected accounts.
\item 1,855,945 accounts were referenced in the list of followings of other accounts, but the API lookup did not return any profile information for these referenced accounts.
Then we tried to perform further API lookups for these referenced accounts, and we obtained profile information for only 137,899 (7.43\%) of them.
For the rest, the API lookups did not return any profile information.
We guess that these accounts were either deactivated~\cite{twitterDeactivateAccount} during the crawl or suspended by Twitter because these accounts violated Twitter's terms of use.
Users can reactivate their account at any time during 30 days after deactivation, so we guess that the observed 7.43\%  have reactivated their accounts.
\item For 5,938 accounts, we did not crawl the list of followings because the API consistently returned an error code.
We counted the number of followings for such accounts as 0.
\item 1,180 user accounts were lost because our archives with data were partially corrupted due to a system bug on two crawling machines.
\end{enumerate}

The number of followings and followers for each account can be obtained in two ways.
Either we get these values from an API call, or we compute them based only on the list of followings for each account.
We use the latter to build our social graph, so we cross-validated the number of following and followers using the latter method with the former one.
We see in Figure~\ref{fig:diff} that there is no difference between the numbers of followers (resp.\ followings) returned by the API and the number of followers (resp.\ followings) in the social graph we computed for 69.14\% (resp.\ 78.79\%) of the collected accounts.
The difference observed for the other accounts is due to three different reasons.
First, our graph does not include protected accounts and their incoming and outgoing arcs, so the number of following and followers in the computed graph is smaller than from the API, which explains that we observe a higher number of positive differences in Figure~\ref{fig:diff}. Second, there is a delay between the time the account information was crawled and the time the list of followings was crawled because of the implementation of the crawler described in Section~\ref{sec:methodology-crawl}.
This delay of 9 hours on average (9.5 minutes median) causes a difference in the number of followings reported by the API and the number of followings obtained by computing the social graph, because some arcs might be added or removed during this delay.
Third, we crawled all accounts during a four months period.
So a given account crawled at time $T$ might be followed (resp.\ unfollowed) by accounts after time $T$, accounts that we crawled after they added (resp.\ removed) the follow links.
Thus, there is a larger (resp.\ smaller) number of followers for this given account in the computed social graph than returned by the API.

\subsection{Measured Twitter Social Graph}
\label{sec:measured-twitter-graph}
We collected all Twitter accounts, consisting of 537~million accounts at the end date of our crawl in July 2012, and accounts' public information (including account creation date, number of published tweets, number of followings and followers, etc.)
We remind that there are 5.97\% of all accounts (32 million) that are protected, which means one needs their approval to get the lists of their followings.
So we collected the list of all followings for non-protected accounts only, resulting in a social graph with 505 million nodes and 23 billion arcs.
The average node in-degree of this graph is 45.6, the median is 1, and the 90th percentile is 33.

Our dataset is, to the best of our knowledge, the largest and most complete dataset of a social network available today.
We also believe that it will be harder in the future to collect such a large and complete dataset.
Indeed, companies are taking measures to prevent large crawls of their social networks.
For instance, Twitter is no more whitelisting machines.
Moreover it has discontinued on June 11, 2013 the API 1.0 that supported anonymous requests and use of already whitelisted machines.
The new API 1.1 requires user authentication for each request making crawls harder and longer to perform.
For these reasons, we acknowledge that our dataset has value to communities interested in social graphs, and we release it for academic use only (with precautions described in Section~\ref{sec:ethical-issues-crawl}) \cite{dataset}.

\subsection{Ethical Issues}
\label{sec:ethical-issues-crawl}
There are two main ethical issues with large scale measurement studies.
First, we need to take care of users privacy.
All data collected in this study are publicly available through the Twitter API, the Twitter applications, and the Twitter Web site.
In particular, we did not collect any data that is not publicly available, or did not work around any protection mechanisms.

Second, we need to respect Twitter terms of use.
We used the regular Twitter API to perform our crawl. We made half of our crawl using machines whitelisted by Twitter, and half of the crawl using a distributed crawler which used the regular Twitter API and conformed to its rate constraint.
On average, we generated from the distributed crawler around 20 requests per second to the API, a rate of requests we believe to be negligible for the Twitter infrastructure.

We release our dataset \cite{dataset} that consists of the Twitter social graph in the format of an adjacency list. In the released dataset each account ID is anonymized.

\section{Graph Analysis Methodology}
\label{analysis_method}
\begin{figure}[t]
\centering
\includegraphics[scale=0.25]{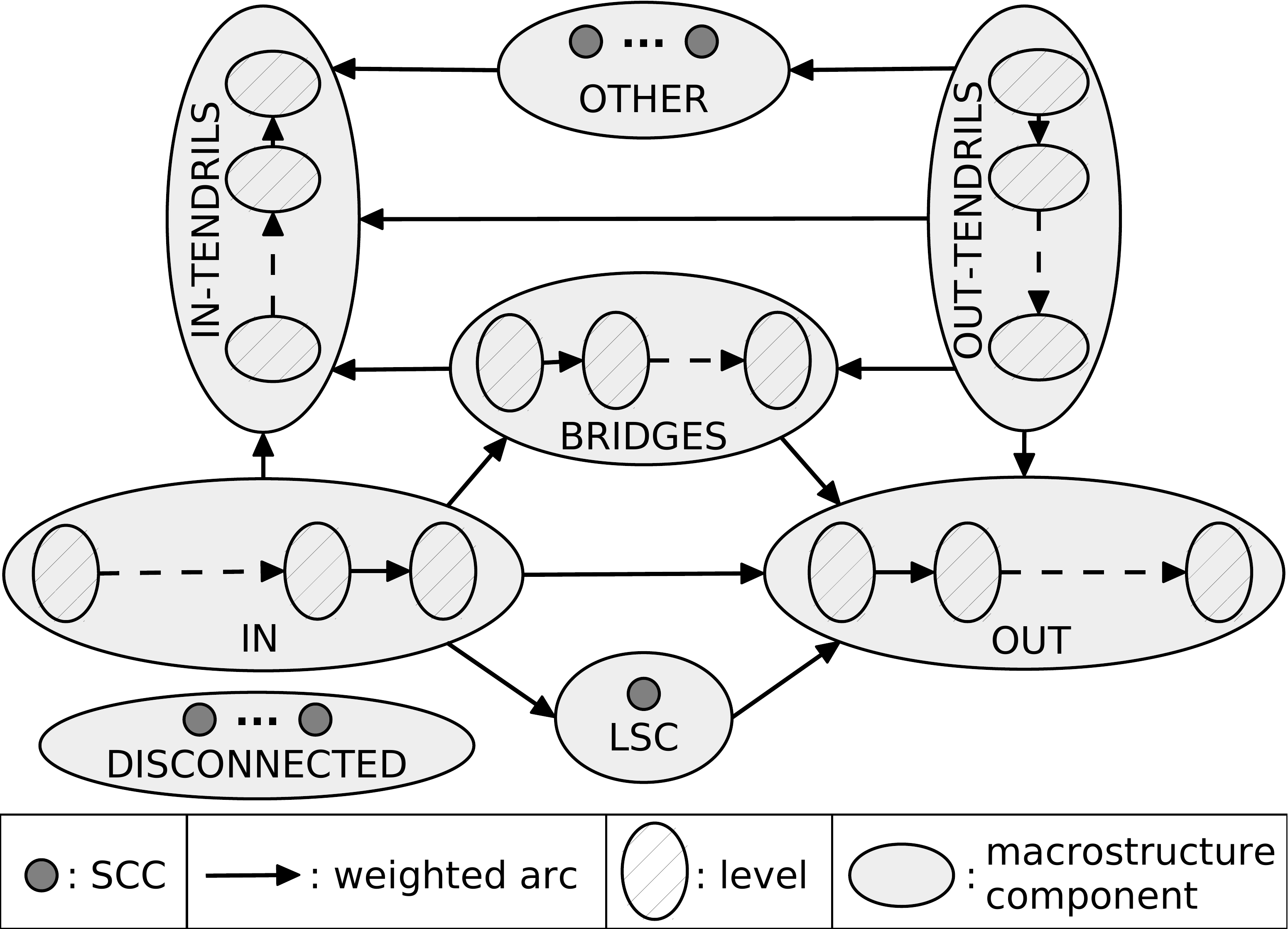}
\caption{
\textbf{Macrostructure of any directed graph.}
}
\label{fig:graph_struct}
\end{figure}

We start discussing the motivation and insights behind the analysis of the macroscopic structure---henceforth called the macrostructure---of the Twitter social graph.
There is a fundamental difference between directed social graphs such as Twitter and other directed graphs such as the Web.
In a directed social graph, not only the links among accounts show the influence of accounts, but they also constrain the propagation of information.
Therefore, unveiling the macrostructure of a social graph sheds light on the highways of information propagation. 

However, it is a challenge to extract a macrostructure on a social graph of the size of Twitter.
The intuition behind our macrostructure analysis is the following.
We want to understand how the Twitter graph constrains the flow of information.
Therefore, we start by identifying all the strongly connected components (SCCs) that are components with a directed path between any two nodes.
In such components, the information can freely circulate, so we abstract each of these components by a single node.
After this stage, we obtain a directed acyclic graph (DAG) that is half of the size of the original graph (in terms of number of nodes), still too large to be analyzed.
Consequently, the next stage is to group nodes in this DAG based on their connectivity to the largest SCC.
As discussed in the following, the largest SCC represents roughly half of the nodes.
This is large and there is undoubtedly an interesting analysis to make on this component, but we keep this analysis for future work and focus in this paper on the macrostructure.
After this stage, we have 8 components representing a tractable graph. We now describe the details of this process.

We compute the macrostructure of the Twitter social graph in two stages.
In the first stage, we use the Tarjan algorithm \cite{tarjan} to compute the SCCs of the Twitter social graph.
Then, we replace each SCC with a single vertex, and the multiple arcs between any two vertices with a weighted arc of weight equal to the number of arcs it replaces.
As a result, we obtain a directed acyclic graph.

In the second stage, to uncover the macrostructure of the directed acyclic graph shown in Figure~\ref{fig:graph_struct}, we use the following procedure.
We first identify the Largest Strongly Connected (\lsc) component, the component with the largest number of original nodes.
From this \lsc{} component, we run a breadth first search (BFS).
We define the set of vertices we find to be the \out{} component, that is the set of nodes with a directed path from the \lsc{} component.
Inside the \out{} component we distinguish \textit{levels} (shown as hatched ellipses on Figure~\ref{fig:graph_struct}).
Each level is a bin of SCCs that have the same distance from the \lsc{} component.
Then we run a reverse BFS from the \lsc{} component and define the set of vertices we find to be the \incomp{} component which is a set of nodes with a directed path to the \lsc{} component.
Similarly to \out{} we distinguish levels inside the \incomp{} component based on the distance to the \lsc{} component.
Next, we perform a BFS starting from the \incomp{} component and a reverse BFS from the \out{} component, reachable nodes that were not yet in the \lsc, \incomp{} or \out{} components were identified as \intend{} and \outtend{} respectively.
Inside the tendrils we can also identify levels depending on the distance to the components these tendrils are growing from.
We separated nodes that were identified as both \intend{} and \outtend{} into the \bridges{} category that consist of accounts connecting the \incomp{} and \out{} bypassing the \lsc{} component, we can also distinguish levels based on the distance to \out{} and distance to \incomp.
After that we put the nodes that were not categorized on previous steps into the \other{} category when there is an undirected path from them to categorized nodes or to \disc{} category otherwise.
All the possible arcs between the components of the macrostructure are shown on Figure~\ref{fig:graph_struct}.

The methodology we describe and the macrostructure representation is inspired from the work of Broder \textit{et al.}~\cite{Broder2000} in the context of the Web for 203 million Web pages.
However, our methodology is significantly different from the one presented by Broder \textit{et al.} Indeed, unlike our methodology that is exhaustive, 
they used a small random sample of 570 nodes from the \lsc{} component to find other components.
This difference in methodology has two important consequences.
First, we perform a complete and accurate classification of all accounts, which is not possible with the methodology of Broder \textit{et al.}, a methodology only intended to show the macrostructure, but not to accurately classify  accounts.
Second, the macrostructure we describe is more detailed and accurate.
In particular, unlike Broder \textit{et al.}, we identified a new component called \other{}, the structure of levels within components, links between components, and the exact number of such links.

In addition, insight we can get from unveiling the macrostructure of the Web is very different from the insight we can get from unveiling the one of a directed social graph such as Twitter.
Indeed, the Web forms a directed graph and the arcs among Web pages are hypertext links.
Therefore, the directed graph of the Web represents the paths to access  Web pages, but no information propagates along the arcs of the graph.
On the contrary, the directed graph of Twitter consists of the follow relationship among accounts.
Each tweet published can only propagate along the paths of this graph. Therefore, whereas the notion of content propagation is irrelevant in the context of the Web graph, it is central in the context of the Twitter graph. 

In summary, we present a method to compute the macrostructure of any directed graph.
Figure~\ref{fig:graph_struct} is not specific to Twitter and can be applied to any directed graph, and in particular to social graphs, where the components group together accounts with different roles in the social graph.
This representation is, to the best of our knowledge, the first attempt to extract \textit{exhaustively} a macrostructure of a large social graph, such as the one of Twitter, taking into account the connectivity of accounts in this graph.
In Section~\ref{twitter2012}, we will discuss the role of the Twitter accounts, depending on the component they belong too.
 
\section{The Macrostructure of Twitter in July 2012}
\label{twitter2012}

\begin{figure}[t]
\centering
\includegraphics[scale=0.25]{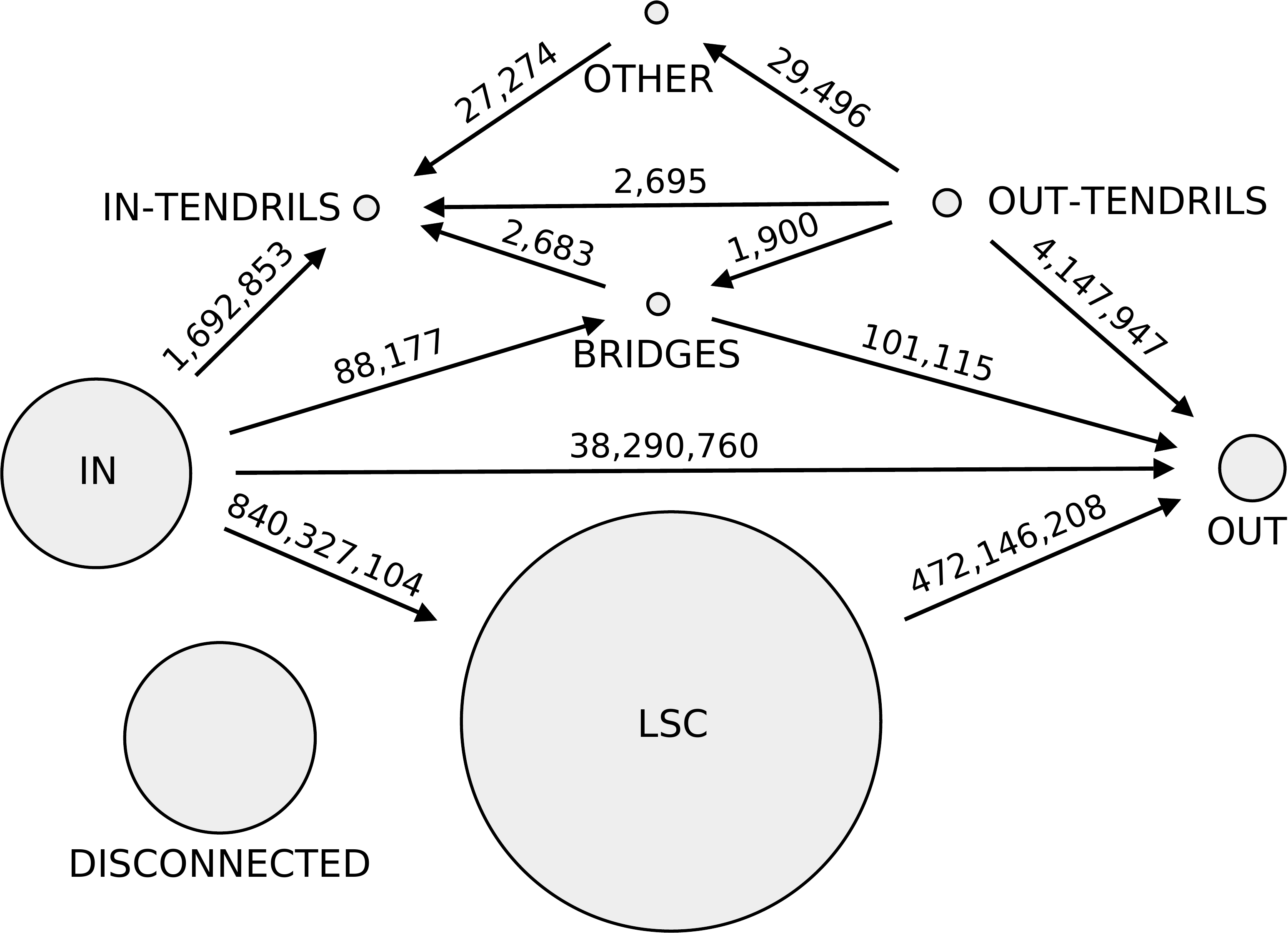}
\caption{
\textbf{Macrostructure of Twitter in July 2012.}
\textnormal{\sl The size of the circles is proportional to the number of accounts in components.
The labels on arrows give the number of arcs between components.}
}
\label{fig:twitter_struct}
\end{figure}

Exploring the macrostructure of the Twitter social graph is interesting because it sheds light on how information propagation is constrained. 
However, this macrostructure would be even more interesting if we can map specific usages of Twitter to components in this macrostructure.
Unraveling a correlation between accounts usages and the macrostructure will improve the understanding of how Twitter is used.

In this section, we dissect the macrostructure of the Twitter social graph, focusing on regular, abandoned, and suspicious accounts.
i) {\bf Regular accounts} are by definition accounts that are neither abandoned or suspended.
Such accounts show the regular activity on Twitter.
ii) {\bf Abandoned accounts} are accounts with few followers and followings, and no recent tweet activity.
Such accounts are important to understand Twitter adoption and to accurately quantify the bias when analyzing Twitter, bias due to these accounts that do not take part in any social activity.
iii) {\bf Suspicious accounts} are often suspended by Twitter because they infringed its terms of use.
We checked that most suspended accounts show evident signs of malicious activity (bunch of sequentially generated accounts, accounts' user name generated with automatic patterns, etc.).
There is no ground truth for the malicious activity, but the notion of suspended accounts is a reasonable metric to detect (in retrospect) malicious accounts~\cite{thomas_suspended_2011}. For the purposes of our study we have recrawled a set of 1~million random users from our dataset on May 6, 2013 to check if they are still active.
In the rest of the paper, we refer to the number of suspended accounts as the number of accounts for which Twitter returned the `suspended' status during this recrawl.

Figure~\ref{fig:twitter_struct} shows the macrostructure of Twitter computed with the methodology presented in Section~\ref{analysis_method}.
We identify 8 components in this graph, with 4 of them (\lsc, \out, \incomp, \disc) representing 98.96\% of all Twitter accounts; so we focus on them.

\begin{table}[t]
\centering
\begin{tabular}{|l|r|r|r|r|r|r|}
\hline
\multicolumn{1}{|c|}{\bf \rotatebox{90}{Component}}& \multicolumn{1}{|c|}{ \bf \rotatebox{90}{Top followed (\%)}} & \multicolumn{1}{|c|}{\bf\rotatebox{90}{Top following (\%)}} & \multicolumn{1}{|c|}{\bf \rotatebox{90}{Top tweeting (\%)}} & \multicolumn{1}{|c|}{\bf \rotatebox{90}{Experts (\%)}} & \multicolumn{1}{|c|}{\bf \rotatebox{90}{Verified (\%)}} &
\multicolumn{1}{|c|}{\bf \rotatebox{90}{Suspended (\%)}} \\
\hline
\lsc & 96.95 & 100 & 88.66 & 94.28 & 97.01 & 1.17\\
\out & 3.05 & 0 & 10.79 & 1.33 & 2.99 & 0.43 \\
\incomp & 0 & 0 &  0.07 & 0.01 & 0 & 1.77\\
DISC. & 0 & 0 & 0.47 & 0.01 & 0 &  5.11 \\
OUT-T. & 0 & 0 & 0 & 0 & 0 &  0.18 \\
IN-T. & 0 & 0 & 0 & 0 & 0 &  0.49 \\
BRID. & 0 & 0 & 0 & 0 & 0 &  0 \\
\other & 0 & 0 & 0.01 & 0 & 0 &  1.25 \\
\hline
\end{tabular}
\caption{
\textbf{Distribution of noteworthy accounts among components.}
\textnormal{\sl The first three columns represent the 10,000 accounts with the largest number of followers, followings, and tweets for the entire Twitter social graph.
The fourth column represents the 2.91 million experts identified by Sharma \textit{et al.}~\protect\cite{sharma_inferring_2012} as influential users in their field (the sum of this column is not 100\% because 4.37\% of the experts are not present in our dataset, most likely because they closed their account, or have been suspended).
The fifth column represents the accounts verified  by Twitter.
The last column represents the percentage of suspended accounts.}
} 
\label{tab:interesting_users}
\end{table} 

\begin{table}[t]
\centering
\begin{tabular}{|l|r|r|r|r|}
\hline
~ &  \multicolumn{2}{|c|}{\bf \begin{tabular}[x]{@{}c@{}}Arcs\\(\%)\end{tabular}} & \multicolumn{1}{|c|}{\bf  \begin{tabular}[x]{@{}c@{}}Tweets\\(\%)\end{tabular}}& \multicolumn{1}{|c|}{ \bf \begin{tabular}[x]{@{}c@{}}Accounts\\(\%)\end{tabular}} \\
\hline
~ & \small followers & \small followings & ~ & ~ \\
\lsc &  98.01 & 96.13 &  98.05 &  50.71 \\
\out & 1.96 & 0.02 & 1.49 & 5.30\\
\incomp & 0.02 & 3.83 & 0.25  & 21.36 \\
DISC. & \textless0.01 & \textless0.01 & 0.21 & 21.60 \\
Others & \textless0.01& 0.02 & \textless0.01 & 1.03 \\
\hline
\bf Total & \multicolumn{2}{|c|}{\bf $23\times10^9 $
} & \bf $127\times10^9 $
& \bf $505\times10^6 $
\\
\hline
\end{tabular}
\caption{
\textbf{Distribution of the arcs, tweets and accounts per component.}
\textnormal{\sl At the scale of the entire Twitter social graph, there is the same number of followings and followers, because they represent the same notion of arc.
But, for each component, the number of followings and followers might be different due to the ingress and egress arcs, so we make a distinction between followings and followers for each component.}
}
\label{tab:dist_arcs_tweets_accounts_per_comp}
\end{table}

The \lsc{} (Largest Strongly Connected) component is the core of the regular Twitter activity.
Indeed, according to Table~\ref{tab:interesting_users}, the \lsc{} component contains 96.95\% of the 10,000 most followed accounts, 100\% of the 10,000 accounts that follow the most, 88.66\% of the 10,000 accounts that tweet the most, 94.28\% of the 2.91 million experts identified by Sharma \textit{et al.}~\cite{sharma_inferring_2012} as influential accounts in their field, and 97.01\% of the verified accounts~\cite{twitterVerifierAccounts} that are accounts of highly sought users (in music, acting, politics, etc.) that Twitter verified to be authentic.
In addition, Table~\ref{tab:dist_arcs_tweets_accounts_per_comp} shows that more than 96\% of the following and follower links, and 98.05\% of the tweets are for accounts in the \lsc{}.

\begin{table}[t]
\centering
\begin{tabular}{|l|r|r|r|}
\cline{1-4}
{\bf Component} & \bf \begin{tabular}[x]{@{}c@{}c@{}}No\\follower\\(\%)\end{tabular} &  \bf \begin{tabular}[x]{@{}c@{}c@{}}No\\following\\(\%)\end{tabular} &  \bf \begin{tabular}[x]{@{}c@{}c@{}}No\\tweet\\(\%)\end{tabular}\\ \cline{2-4}
\cline{1-4}
\lsc & 0  & 0  & 23.87\\
\out & 0  & 92.97  & 61.82\\
\incomp  & 96.13  & 0  & 60.10\\
\disc  & 99.63  & 99.63  & 79.31\\
\outtend & 99.13 & 0  & 73.20\\
\intend & 0 & 98.78  & 70.40\\
\bridges  & 0 & 0 & 67.34\\
\other & 51.39 & 46.67 & 67.56\\
\cline{1-4}
\hline
\end{tabular}
\caption{
\textbf{Percentage of accounts with no follower, no following or no tweet per component.}
}
\label{tab:no}
\end{table} 

However, it is wrong to believe that the \lsc{} component is the only one that matters when studying Twitter, other components contain a lot of accounts with specific roles in the Twitter ecosystem.
We see in Table~\ref{tab:dist_arcs_tweets_accounts_per_comp} that the \lsc{} contains only 50.71\% of all accounts.
This is surprising because it is easy to be part of the \lsc{} component, an account only needs one following and one follower already in the \lsc{} component.
Also, we observe that a large fraction of the suspicious activity in Twitter is outside of the \lsc{} component, as we see in Table~\ref{tab:interesting_users} (last column).
Finally, when looking at the percentage of accounts with no follower, no following, or no tweet, we see in Table~\ref{tab:no} that each of the four main components has fundamentally different characteristics.
Indeed 92.97\% of the accounts in the \out{} component have no following, 96.13\% of the accounts in the \incomp{} component have no follower, and almost all accounts in the \disc{} component have no following and no follower.
Moreover, at least 60\% of the accounts in these three components never sent any tweet, whereas it is only 23.87\% for the \lsc{}.

In summary, we see that even if most of the regular Twitter activity is in the \lsc{} component, other components contain half of the Twitter accounts and present characteristics worth studying.
In the following, we dig into each component to discuss its main characteristics.

\subsection{\lsc{} Component}
\label{sec:lsc-comp}

\begin{figure}[t]
\centering
\includegraphics[scale =0.47]{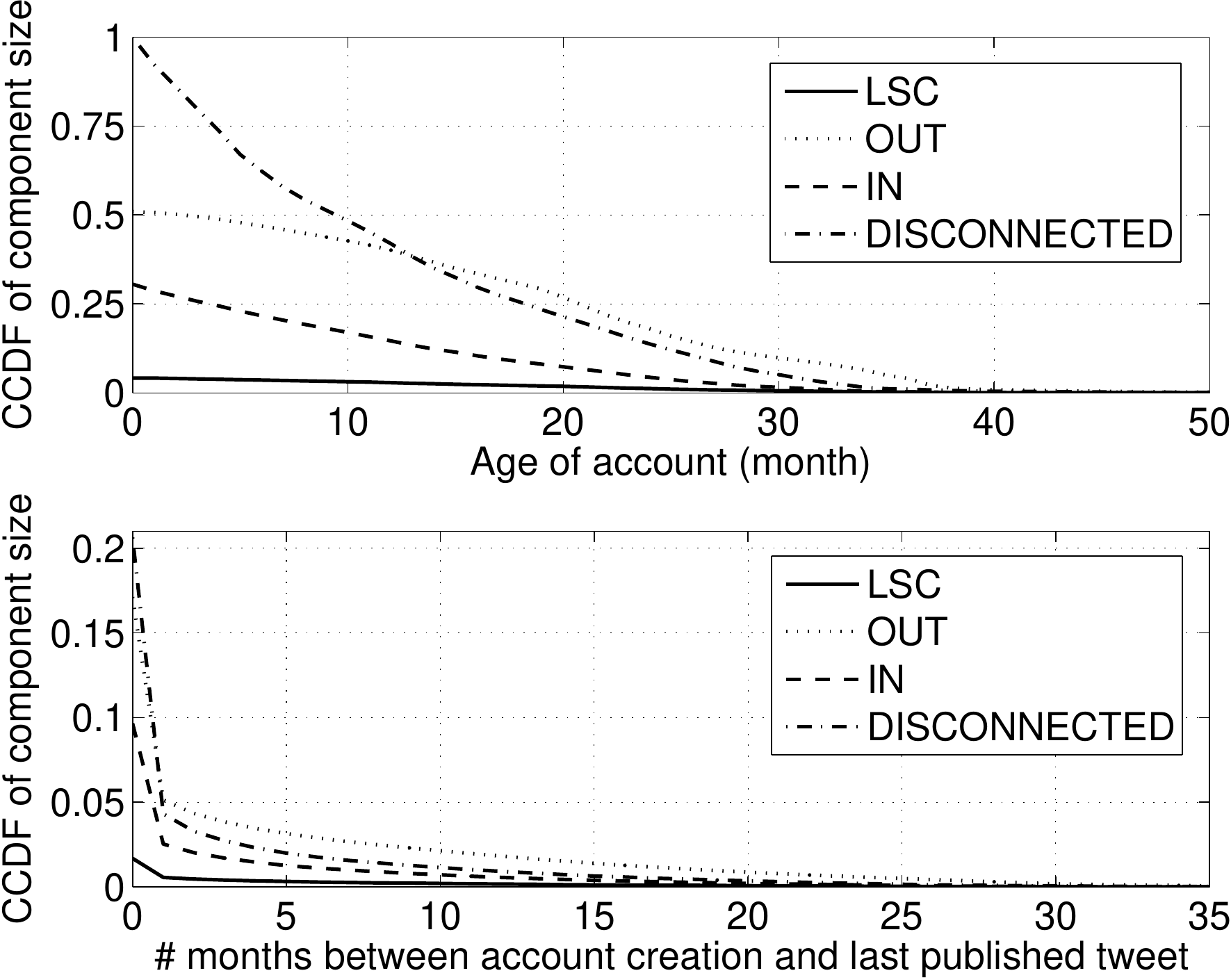}
\caption{
\textbf{Characterization of abandoned accounts.}
\textnormal{\sl (top) Identification of old abandoned accounts.
CCDF of accounts with at most one follower and one following in a component according to the account creation date.
(bottom) Characterization of accounts who published at least one tweet.
CCDF of the duration between the creation date of an account and the date of its last published tweet for accounts with at most one follower and one following.}
}
\label{fig:abandoned-dist-by-cat}
\end{figure}

We have seen that most of the regular Twitter activity is in the \lsc{} component.
However, due to the simplicity to belong to the \lsc{} component, many abandoned and suspicious accounts also belong to it.

\subsubsection{Abandoned Accounts}
\label{sec:lsc-abandoned}
Most accounts with one following and one follower in the \lsc{} are abandoned accounts.
We see in Figure~\ref{fig:abandoned-dist-by-cat} (top, solid line) that there are 4.18\% of accounts in the \lsc{} component with one following and one follower.
In addition, out of the accounts with one following and one follower in the \lsc{} component, 86.34\% are more than 6 months old and 59.57\% never sent any tweet.

In summary, a large fraction of accounts in the \lsc{} component with one following and one follower did not have any change in their number of followings and followers for months and did not send tweets recently.
Considering that it is unlikely that such accounts will actively follow a single other account for month (so no serious follow activity) without tweeting anything (so no publishing activity), it is reasonable to believe that these accounts are abandoned.

\begin{table}[t]
\centering
\begin{tabular}{|l|r|r|r|r|r|}
\hline
\multicolumn{1}{|c|}{\bf \rotatebox{90}{Component}} & \multicolumn{1}{|c|}{\bf  \rotatebox{90}{Top followed (\%)}} & \multicolumn{1}{|c|}{\bf  \rotatebox{90}{Top following (\%)}} & \multicolumn{1}{|c|}{\bf  \rotatebox{90}{Top tweeting (\%)}}& \multicolumn{1}{|c|}{\bf  \rotatebox{90}{\begin{tabular}[x]{@{}c@{}}Top following with\\ $\le1$ follower (\%)\end{tabular}}} &  \multicolumn{1}{|c|}{\bf  \rotatebox{90}{\begin{tabular}[x]{@{}c@{}}Top tweet with\\ $\le1$ follower (\%)\end{tabular}}}\\
\hline
\lsc &0.33 & 1.15 & 1.99 & 97.83 & 3.02 \\
\out & 1.15 & 10.30 & 5.20 & 0.45 & 5.26 \\
\incomp & 2.78 & 96.87 & 3.87 & 96.87 & 3.89 \\
DISC. & 1.38 & 1.33 & 7.43 & 2.84 & 7.48\\
\hline
\end{tabular}
\caption{
\textbf{Percentage of suspended accounts (on the 6th of May 2013) per component for 5 outlier categories.}
\textnormal{ \sl The first three columns represent the 10,000 accounts with the largest number of followers, followings, and tweets for the entire Twitter social graph.
The fourth column is for the 10,000 accounts with the largest number of followings and at most one follower.
The last column is for the 10,000 accounts with the largest number of tweets and at most one follower. }
}
\label{tab:suspended_users}
\end{table} 

\begin{figure}[t]
\centering
\includegraphics[scale = 0.47]{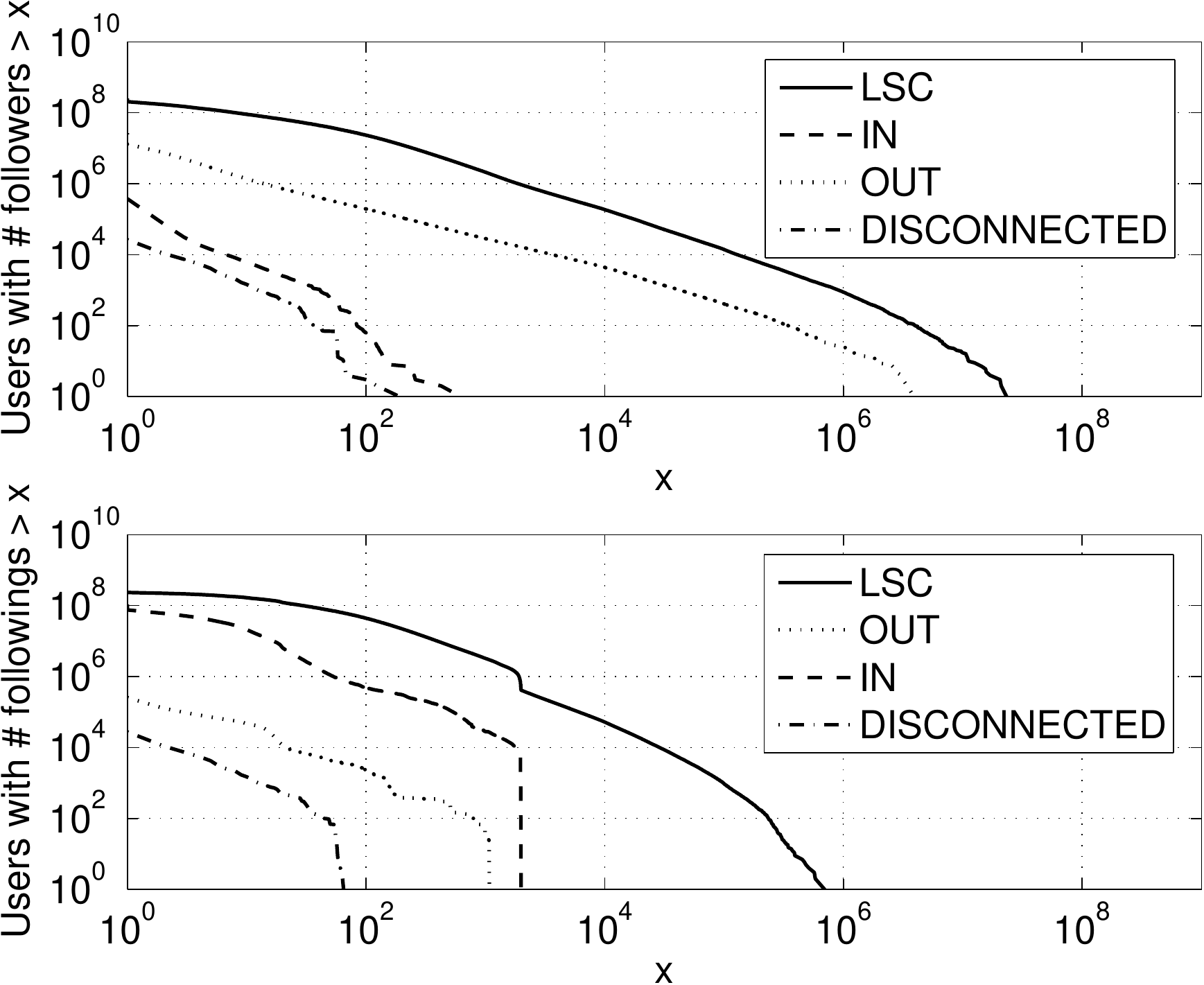} 
\caption{
\textbf{Distribution of followers (top) and followings (bottom) by category.}
\textnormal{\sl Accounts with no follower (top) and no following (bottom) are filtered out (see Table~\ref{tab:no})}
}
\label{fig:revert-CDF-followings-followers}
\end{figure}
\eject
\subsubsection{Suspicious Accounts}
\label{sec:lsc-suspicious}
The \lsc{} component also contains suspicious accounts.
We present in Table~\ref{tab:suspended_users} the percentage of suspended accounts per component for five outlier categories. An \emph{outlier} account is followed, following, or tweeting much more than a regular account, thus it is a good candidate for suspicious activity. 
The first three columns represent accounts with the largest number of followers, largest number of followings, and largest number of tweets.
The fourth and fifth columns are for the accounts with the largest number of followings and tweets, but with at most one follower.
We consider this notion of outliers because following a lot of accounts is a known technique used by spammers \cite{thomas_suspended_2011}.
To reduce the impact of spammers, we remind that Twitter imposes a limit of 2,000 followings for accounts with no follower, and then a linear increase with the number of followers.
Accounts close to this limit and with at most one follower are more likely to be spammers.
The last column is for accounts that send the largest number of tweets, but with at most one follower.
This is also a suspicious behavior, because it is strange to send a lot of tweets if nobody (or a single other account) follows them.
Spammers can send a lot of tweets to interfere with trending topics or the Twitter search functionality, and to direct messages to a specific user using \textit{@mentions}~\cite{thomas_suspended_2011}.

Considering the huge number of suspicious accounts, we cannot afford to manually inspect all of them.
Therefore, we consider a suspicious account to be malicious if it was suspended by Twitter, see Table~\ref{tab:suspended_users}. 

As expected, the top followed accounts in the \lsc{} component are regular, only 0.33\% have been suspended.
Indeed, it is complex to manipulate the number of followers, because it requires to either manipulate other accounts in order to incite them to follow, or to create fake accounts whose only one goal is to follow.
More surprising, the top following accounts are also regular for Twitter, only 1.15\% have been suspended.
We expect accounts that follow a lot of other accounts to be spammers, but, according to Figure~\ref{fig:revert-CDF-followings-followers} (bottom), the \lsc{} component is the only one to have accounts that break the limit of 2,000 followings.
So the top following in the \lsc{} component also have a lot of followers, thus the low number of suspended accounts.

Then we observe in Table~\ref{tab:suspended_users} two important behaviors that characterize well the outlier activity in the \lsc{} component.
First, 97.83\% of the top following with at most 1 follower have been suspended.
This means that most of the accounts in the \lsc{} component close to the limit of 2,000 followings are malicious.
Second, only 3.02\% of the top tweeting accounts, but with at most a single follower have been suspended.
The rest looks like regular for Twitter.
By manually inspecting these accounts that looks regular for Twitter, we found bots used as an interface to job forums, news site, Yahoo!Answers, YouTube published videos, etc.
So, it seems that Twitter is used by developers to generate a stream of data collected from third party Web sites.
As these accounts have only one follower, we guess that they are either used for tests only, or that the developers are using a Twitter widget to embed their account timeline into a Web site.

\subsection{\out{} Component}
\label{sec:out-comp}

\begin{figure}[t]
\centering
\includegraphics[scale =0.47]{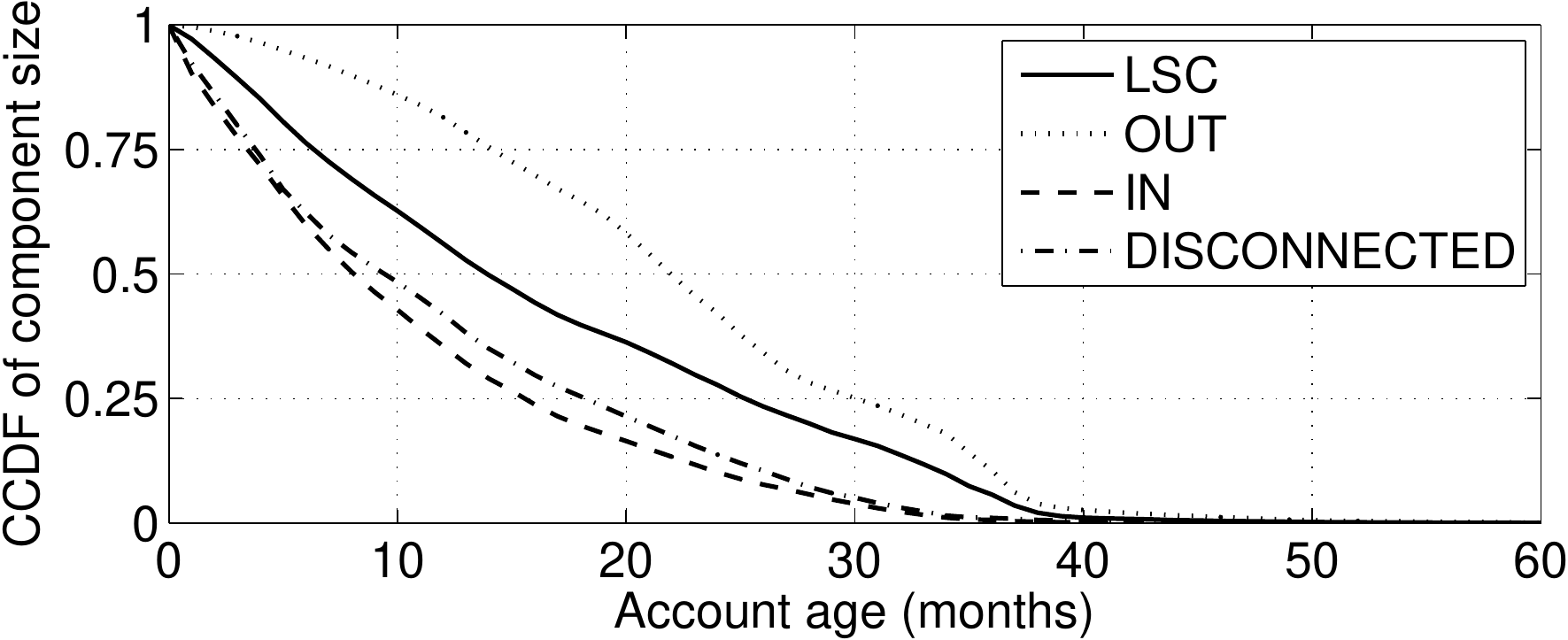}
\caption{
\textbf{Age of accounts in each component.}
\textnormal{\sl CCDF of accounts in a component according to the account creation date.}
}

\label{fig:age-per-comp}
\end{figure}

The \out{} component represents all Twitter accounts with a directed path from the \lsc{} component.
In addition, these accounts can also have directed paths from other components, but no account in \out{} can have a directed path to any other component (directed paths among \out{} accounts are possible, so if an \out{} account has following links, they necessarily come to other \out{} accounts).

\subsubsection{Regular Accounts}
\label{sec:out-regular}
A specificity of the \out{} component is that a small set of accounts (belonging to celebrities) attract most of the follower links for this component.
These are regular \out{} accounts.
We see in Figure~\ref{fig:twitter_struct} that more than 500 million links between components are directed to \out, 37.93\% of all inter-components links, whereas the \out{} component represents only 5.30\% of all accounts.
Also, we see in Table~\ref{tab:dist_arcs_tweets_accounts_per_comp} that accounts in \out{} presents 1.96\% of all follower links, which make it the second component with the largest number of follower links (we sum all follower links for all accounts in a given component).
Among the 100 accounts that have the largest number of followers, we found that there are 35 verified accounts representing 12\% of the arcs from the \lsc{} to \out.
These accounts are owned by celebrities that belong to the \out{} component because they do not follow any other account.

We observe another interesting specificity of the \out{} component in Figures~\ref{fig:abandoned-dist-by-cat} (top) and Figures~\ref{fig:age-per-comp}.
The \out{} component is the only one to show an inflection point for both curves around 20 months, meaning that the proportion of recent accounts in the component is lower than for other components.
To explain this inflection point, we need to characterize the kind of accounts that stay in the \out{} component.
According to Table~\ref{tab:no}, 92,97\% of the \out{} accounts have no followings, but they all have at least one follower because they belong to the \out{} component.
These accounts are what we call selfish (they are not interested in tweets from other accounts), a decreasing trend in Twitter in the past two years. We will discuss further this trend in Section~\ref{sec:repart-new-acco}.

\subsubsection{Abandoned Accounts}
\label{sec:out-abandoned}
As we discussed in Section~\ref{sec:lsc-comp}, most accounts with at most one following and one follower are also abandoned accounts for the \out{} component.
We see in Figure~\ref{fig:abandoned-dist-by-cat} (top) that 50.94\% of the accounts have at most one following and one follower, and 40.11\% are more than 1 year old.
We see in Figure~\ref{fig:abandoned-dist-by-cat} (bottom) that 82.89\% of the accounts with at most one following and one follower never sent a tweet, and that only 5.13\% of the accounts with at least one following and one follower sent a tweet more that 1 month after their creation date.
This is a consequence of the \textit{Find friends} feature available in Twitter that allows users to search their entire contact lists for Twitter accounts.
By default, once the search is done, all accounts are checked to be followed.
As a consequence, we observe many accounts in the \lsc{} component that followed abandoned accounts in the \disc{} component, making these abandoned accounts move to the \out{} component.

\subsubsection{Suspicious Accounts}
\label{sec:out-suspicious}
There are fewer malicious accounts in the \out{} component than for other components.
We see in Table~\ref{tab:suspended_users} that the percentage of suspended accounts for outlier accounts is low for the \out{} component.
We explain the low number of suspended accounts for the top followings because no account reaches the limit of 2,000 followings, and that few accounts have more than a hundred followings, see Figure~\ref{fig:revert-CDF-followings-followers} (bottom).
As long as an account in \out{} follows an account in the \lsc{}, it belongs to the \lsc{}, so spammers using following links to spam are likely to escape in the \lsc{} component.
We explain the low number of suspended accounts for the top tweeting with at most one follower because (as for the \lsc{} component) most of these accounts are operated by bots.
Finally, we see in Table~\ref{tab:interesting_users} that out of the 4 main components, \out{} is the component with the smallest number of malicious accounts.

\begin{figure}[t]
\centering
\includegraphics[scale = 0.47]{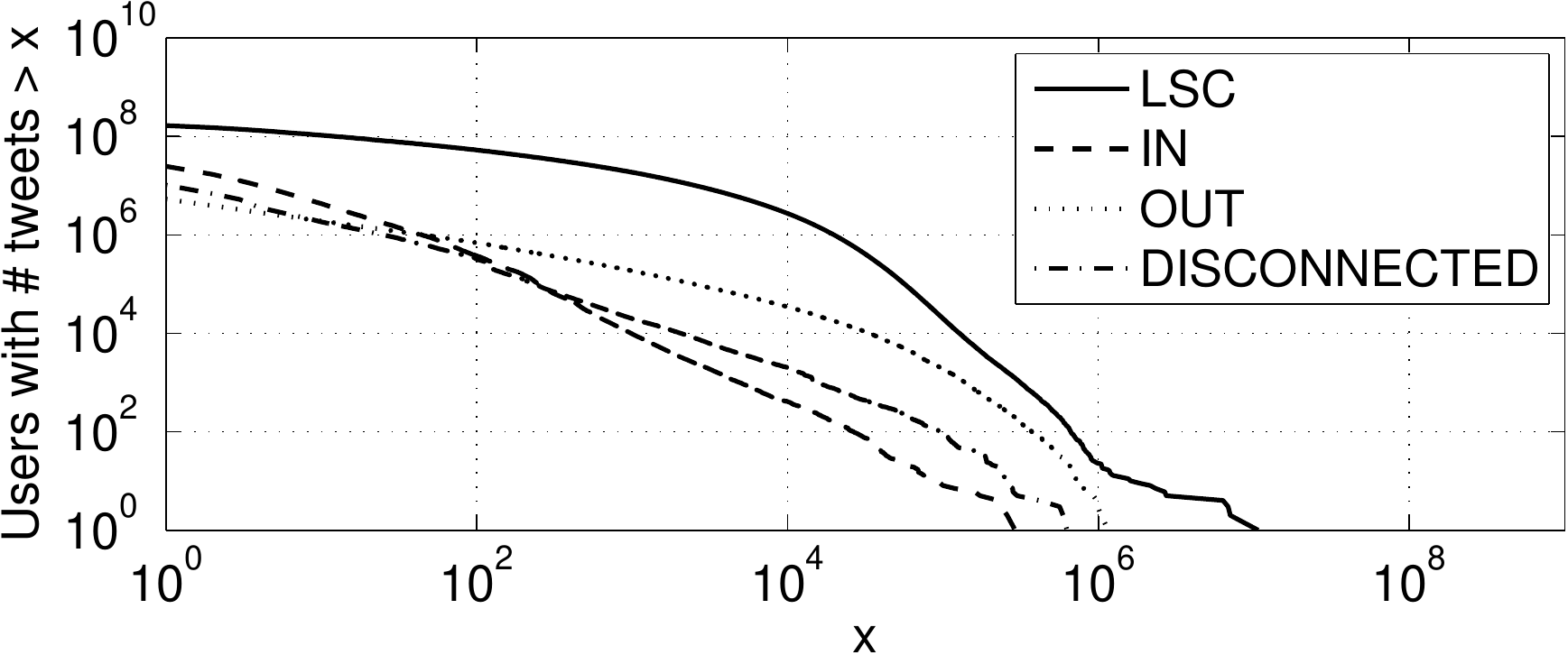}
\caption{
\textbf{The distribution of number of tweets by component.}
\textnormal{\sl Accounts  with no tweets are filtered out (see Table~\ref{tab:no}).}
}
\label{fig:tweets-main}
\end{figure}

\subsection{\incomp{} Component}
\label{sec:in-comp}
The \incomp{} component is much different from the two previous ones because accounts in this component have few followers (see Figure~\ref{fig:revert-CDF-followings-followers}, top) and the distribution of the number of tweets is very different (see Figure~\ref{fig:tweets-main}) from the ones of the \lsc{} and \out{} components.
The \incomp{} component contains the second largest fraction of abandoned and suspicious accounts, after the \disc{} component.

\subsubsection{Regular Accounts}
\label{in-regular}
The regular users for the \incomp{} component are passive followers, that are accounts who follow accounts in the \lsc{}, but never publish tweets and are not followed.\ 
Indeed, in Tables~\ref{tab:dist_arcs_tweets_accounts_per_comp} and \ref{tab:no} we see that the \incomp{} component contains 21.36\% of all Twitter accounts, but 96.13\% of them have no follower, and 60.10\% of them published no tweet (we remind that accounts with followers in \incomp{} are followed by other accounts in \incomp{} only).
This component consists of accounts who follow accounts in the \lsc{} (99.6\%) or an account in \incomp{} (0.4\%).
We will see in Section~\ref{sec:repart-new-acco} that the trend of accounts to be passive followers on Twitter (that is, belong to \incomp{} component) has been growing since 2009.

Many accounts belonging to the \incomp{} component move to the \lsc{} component.
We see in Figure~\ref{fig:abandoned-dist-by-cat} (top) that 30.56\% of the accounts in the \incomp{} component have at most one following and one follower, but that only 14.61\% are more than one year old.
So even if few of them have been tweeting close to the creation date of their accounts (see Figure~\ref{fig:abandoned-dist-by-cat}, bottom), it is likely that they moved to the \lsc{} component and tweeted from it.
Indeed, we see in Table~\ref{tab:interesting_users} that only 1.77\% of the accounts in the \incomp{} component have been suspended, but that accounts are much younger in the \incomp{} component than in the \lsc{} and \out{} components, see Figure~\ref{fig:age-per-comp}.

\subsubsection{Abandoned Accounts}
\label{sec:in-abandoned}
Whereas 96.13\% of the accounts in the \incomp{} component never published any tweet (see Table~\ref{tab:no}), the fraction of abandoned accounts is much lower in this component than in the \out{} and \disc{} components. Indeed, we see in Figure~\ref{fig:abandoned-dist-by-cat} (top) that only 30.56\% of the accounts in the \incomp{} component have at most one following and one follower, and 20.88\%  have at most one following, one follower, and never published any tweet. Moreover, according to Figure~\ref{fig:revert-CDF-followings-followers} (bottom), 23.04\% of the accounts follow at least 10 other accounts, thus a passive follower activity. 

\subsubsection{Suspicious Accounts}
\label{sec:in-suspicious}
The \incomp{} component contains many malicious accounts among the outliers.
We see in Table~\ref{tab:suspended_users} that 96.87\% of the accounts with the largest number of followings are suspended.
We note that all top followings have at most 1 follower in this component.
There is also 3.87\% of the accounts that tweeted the most that were suspended.
For the rest, after manual inspection, we also found, as for the two previous components, that they are used by bots.

Finally, the \incomp{} component has a very interesting property for people looking for a reliable metric to assess influencers.
Cha \textit{et al.}~\cite{ChaEtAl2010} show that the number of followers is not a reliable metric, because users perform link farming \cite{ghosh_understanding_2012} to increase their number of followers. However, this is a rare problem in the \incomp{} component.
Indeed, accounts in the \incomp{} are clearly not interested in increasing their number of followers (see Figure~\ref{fig:revert-CDF-followings-followers}, top) thus the accounts they follow will not be biased by this problem.
Evaluating the benefit of considering accounts in the \incomp{} to assess influencers is an interesting problem for future work.

\subsection{\disc{} Component}
\label{sec:disc-comp}

Accounts in the \disc{} component, like in the \incomp{} one, have few followers (see Figure~\ref{fig:revert-CDF-followings-followers}, top) and the distribution of their number of tweets is very different (see Figure~\ref{fig:tweets-main}) from the ones of the \lsc{} and \out.
The \disc{} component contains the largest fraction of abandoned and suspicious accounts.
There are almost no regular users in this component.

\subsubsection{Abandoned Accounts}
\label{disc-abandoned}
A specificity of the \disc{} component is that is contains a lot of abandoned accounts.
In spite of being the second largest component with 21.6\% of all accounts (see Table~\ref{tab:dist_arcs_tweets_accounts_per_comp}), 78.94\% of accounts in the \disc{} component have no followers and no followings, and never published any tweet.
Furthermore, 72.44\% of accounts in \disc{} component are older than one month.
Therefore, we can conclude that the \disc{} component has, by far, the largest number of abandoned accounts.
We see in Fig.~\ref{fig:abandoned-dist-by-cat} (top) that 99.97\% of its accounts have at most one following and one follower, but only 41.93\% of them are older than 12 months.
Like for the \incomp{} component, many account in the \disc{} component are recent (see Figure~\ref{fig:age-per-comp}), thus some accounts in this component have moved to another component.

\subsubsection{Suspicious Accounts}
\label{sec:disc-suspicious}
Finally, we see in Table~\ref{tab:interesting_users} that the \disc{} component contains the largest fraction of malicious accounts, but we don't observe in Table~\ref{tab:suspended_users} an outlier category grouping them.
Indeed, most accounts have no followings, no followers and no tweets, so the number of outliers is much smaller than our sample size.

In summary, the \disc{} component hosts a lot of abandoned accounts and a large fraction of the malicious activity on Twitter, it is also a transitional place for new accounts before they migrate to another component.

\subsection{Other Components}
\label{sec:other-comp}

The smallest components, \intend, \outtend, \bridges, and \other{} represent 1.03\% of all accounts.
Most accounts in these components are either accounts created for test, or new accounts that will migrate to another component after some time.
We do not discuss deeper these components as their impact on the Twitter social graph is small compared to the 4 main components.

\subsection{Discussion}
\label{sec:results-discussion}
We can draw several important lessons from the results discussed in this section.

First, the macrostructure of the Twitter social graph significantly constrains the propagation of information.
Therefore, models of information propagation in social networks might lead to wrong results when abstracting the underlying social graph.
This work sheds light on how to correctly abstract the social graph, and because the macrostructure is reasonably simple, with 3 main components with active accounts, we believe it is possible to model the underlying graph constraint.

Second, we identify a correlation between components in the macrostructure and the usage of accounts in these components.
This result challenges the sampling techniques that follow arcs (such as random walks or bi-directional breadth first search) because the statistical validity of the sample might be low.
For instance, all sampling techniques following arcs that start from well connected (or active) accounts will miss all of the malicious activity located in the \disc{} component.

Last, the identification of the role of accounts in each components is important to understand who are the influencers in Twitter.
For instance, as discussed in Section~\ref{sec:in-suspicious}, users try to increase their popularity in Twitter by either offering reciprocation to the users that accept to follow them, or by buying follower links on the black market.
Therefore, we can identify real influencers by focusing exclusively on the followers in the \incomp{} component, removing suspicious accounts by filtering out all accounts younger than, e.g., six months.

\section{Evolution of the Macrostructure of the Twitter Social Graph with Time}
\label{evolution}

\begin{figure}[t]
\centering
\includegraphics[scale=0.47]{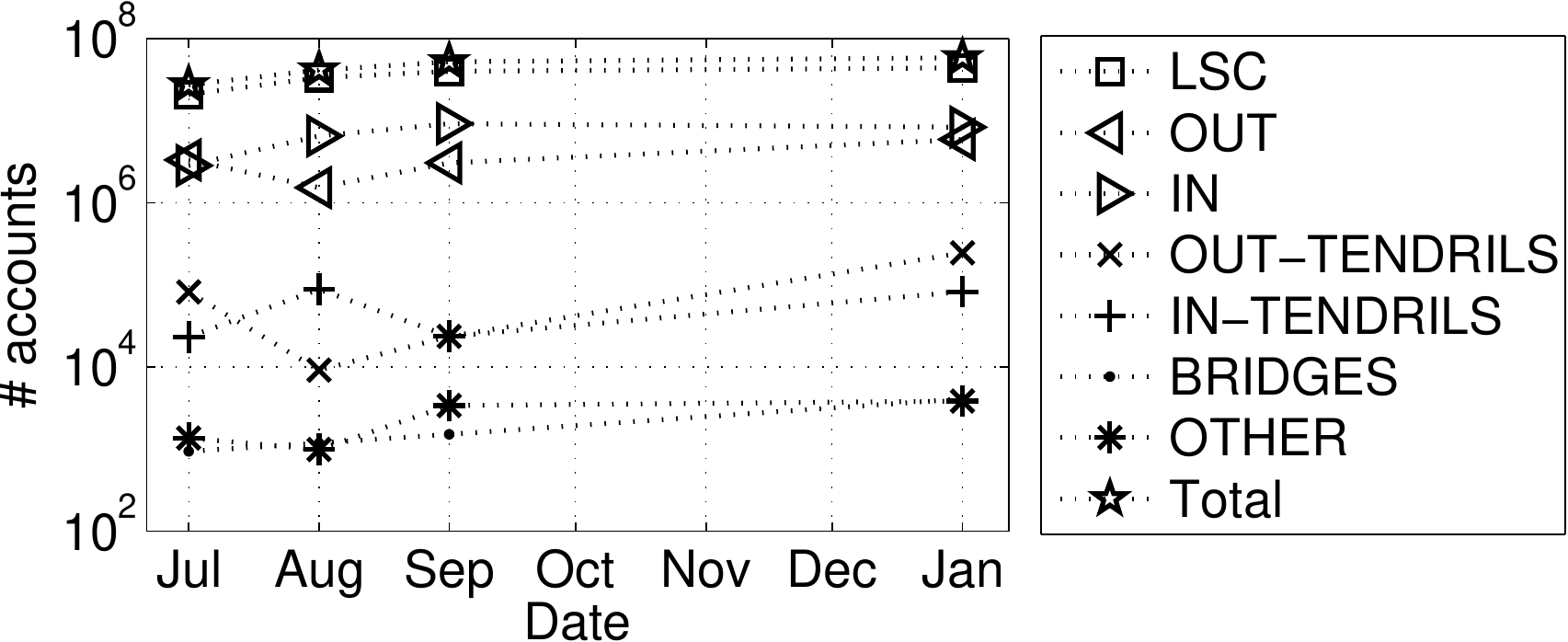}
\caption{
\textbf{Comparison of our estimated graphs of 2009 (labeled Jul and Jan) with two existing Twitter datasets made in August \protect\cite{KwakEtAl2010} and September \protect\cite{ChaEtAl2010} 2009.}
\textnormal{\sl Our simple methodology gives an approximation of the macrostructure of the Twitter social graph that is consistent with existing datasets.}
}
\label{fig:compare2009}
\end{figure}

In this section, we discuss the evolution of the macrostructure of the Twitter social graph with time from January 2007 to July 2012.
To present this evolution, we first describe the estimation technique we use to estimate the Twitter social graph in the past.
Then we validate our technique using two public datasets collected in 2009~\cite{KwakEtAl2010,ChaEtAl2010}.
We discuss the evolution of the macrostructure of the Twitter social graph with time and explain how new accounts led to the time evolution we observed, shedding light on the evolution of the usage of Twitter in the past 6 years.

\subsection{Methodology to Estimate the Macrostructure}
\label{sec:meth-estim-twitt}

The evolution with time of the macrostructure of the Twitter social graph is interesting, because it shows the evolution of the Twitter usage.
We have seen in Section~\ref{twitter2012} that components represent specific categories of usage.
However, the Twitter API does not give access to the past social graph of Twitter. 

We propose a simple approach to approximate the macrostructure of the Twitter social graph.
The dataset we describe in Section~\ref{sec:measured-twitter-graph} covers all Twitter accounts in July 2012 (with the limitation described in Section~\ref{sec:limitations-crawl}), and for each account we have the creation date.
To approximate the macrostructure of the Twitter social graph at date $D$, we remove from our dataset all accounts created after this date, and all arcs to and from these accounts.
Then, we use the methodology described in Section~\ref{analysis_method} to compute the macrostructure of the resulting graph at date $D$. 

This simple methodology has two important limitations.
First, we do not have any suspended and deactivated accounts in our dataset.
Accounts are suspended by Twitter because they infringed the terms of use, most of the time they are spammers.
Deactivated accounts have been closed by users themselves.
We believe such accounts, when they were still active, had a small impact of the Twitter social graph.
Second, the Twitter API does not give access to the arc creation date\footnote{For a given account, Meeder \textit{et al.} observed that the 1.0 Twitter API returned the arcs in an order that was the reverse order of creation of the arcs for this account~\cite{meeder_we_2011}. Our recent experiments with the Twitter API have shown that it is no more possible to rely on this ordering property. 
}.
Therefore, we assume that all arcs between any two accounts in July 2012 existed at date $D$ as long as the two accounts existed at this date; equivalently we assume that if there is an arc between two accounts, it is created close to the creation date of the youngest account.
We are aware that, as reported by Kwak \textit{et al.}, the creation of arcs among accounts is more complex than our simple approximation \cite{kwak_fragile_2011}. However, our goal is to understand the evolution of the macrostructure of the Twitter social graph with time, not the fine grain evolution of arcs between accounts. For this reason, we believe that our approximation is reasonable. 

Moreover, to validate this approximation on creation dates of arcs, we compare our approximation with two datasets collected in 2009 \cite{KwakEtAl2010,ChaEtAl2010}.
Kwak \textit{et al.}~\cite{KwakEtAl2010} and Cha \textit{et al.}~\cite{ChaEtAl2010} independently collected two Twitter datasets in August 2009 and September 2009 respectively and used different methodology. Kwak \textit{et al.} used a technique close to a BFS and reverse BFS from a popular account and also collected accounts referring to trending topics (so active accounts only), and Cha \textit{et al.} used a crawl by account ID (as we did).
For each of the two datasets we computed the Twitter macrostructure according to the methodology described in Section~\ref{analysis_method}, and we approximated the macrostructure of Twitter using our dataset in July 2009 and January 2010.
We show in Figure~\ref{fig:compare2009} the result of this validation: the order of the size of each components is consistent between the two validation datasets and our dataset.
In addition, we have compared the dataset by Kwak \textit{et al.} with our closest estimation (July 2009). We found that 88.25\% of the users common to both datasets belong to the same components in both datasets. We cannot make such a validation for the second dataset because Cha \textit{et al.}\ have anonymized it.
In summary, the dynamics of the creation and deletion of arcs is complex \cite{kwak_fragile_2011}, but we have shown that our simple approximation is reliable enough for the purpose of our macrostructure study.

There is no \disc{} component in Figure~\ref{fig:compare2009}, because this component is missing in the two validation datasets.
Either the methodology did not permit to crawl accounts in this component \cite{KwakEtAl2010}, or these accounts were filtered out in the published dataset \cite{ChaEtAl2010}.
We observe in Figure~\ref{fig:compare2009} some small variations for the \out{}, \intend{}, and \outtend{} components between the two validation datasets and our dataset.
These variations can be explained by a major change in the Twitter macrostructure that happened in 2009.
We discuss further this change in the next section.

\begin{figure}[t]
\centering
\includegraphics[scale =0.47]{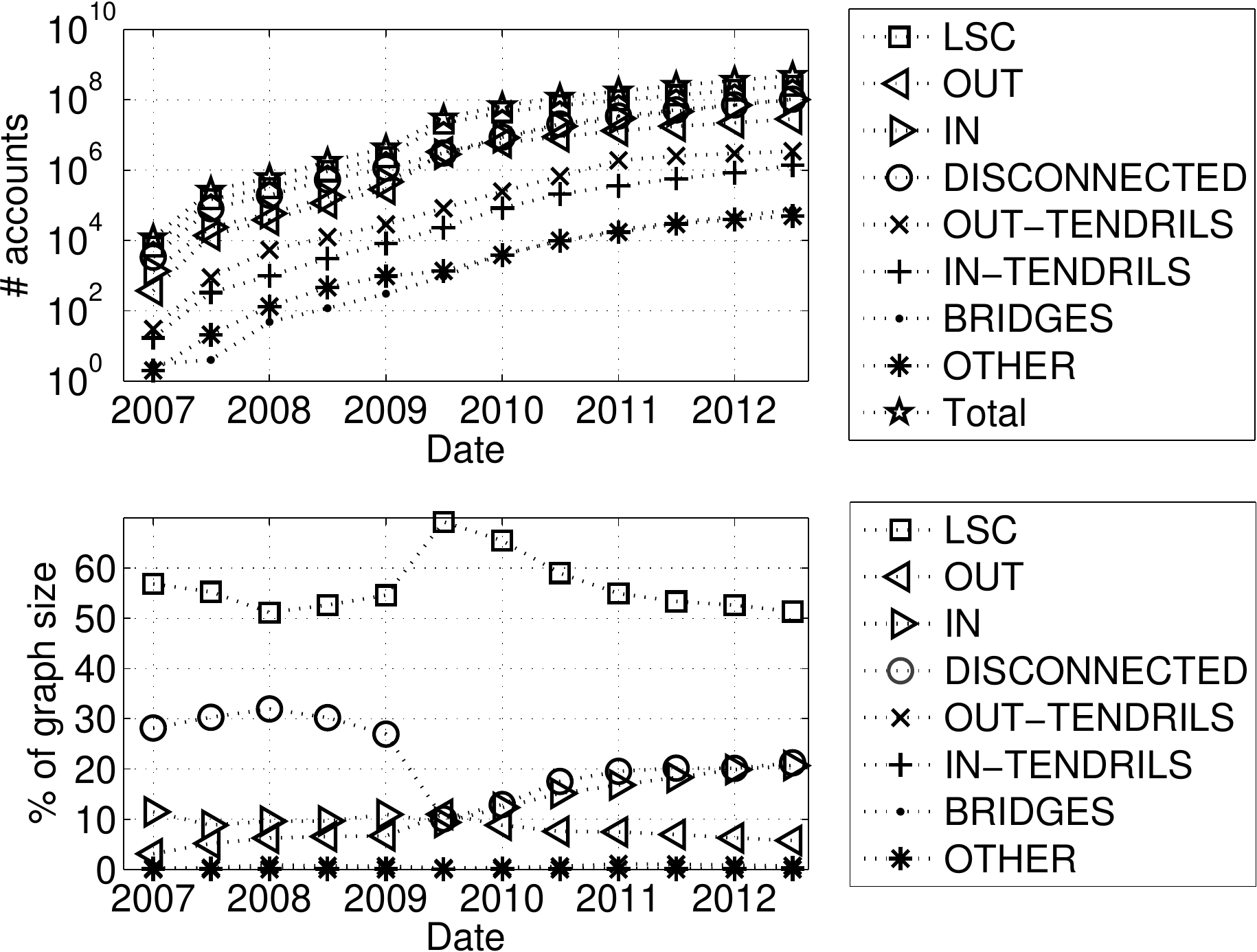}
\caption{
\textbf{The estimated evolution of the macrostructure of Twitter with time.}
\textnormal{\sl (top) Sizes of components in log scale.
(bottom) Sizes of the same components as a percentage of the size of the graph.}
}
\label{fig:evolution}
\end{figure}

\subsection{Evolution of the Macrostructure}
\label{sec:evol-twitt-graph}

To observe the evolution of the Twitter social graph with time, we approximate its macrostructure using the simple methodology discussed in Section~\ref{sec:meth-estim-twitt} every six months from January 1, 2007 to July 1, 2012.
The first account on Twitter was created on March 21, 2006, but due to the small number of accounts created between March and July 2006, we decided to skip the macrostructure of the Twitter social graph in July 2006 and start our analysis in January 2007.

We see in Figure~\ref{fig:evolution} (top) the evolution of the size of each component with time, confirming that the \lsc{}, \out{}, \incomp{}, and \disc{} have always been the largest components in Twitter.
However, by looking at the size of each component normalized with the graph size in Figure~\ref{fig:evolution} (bottom), we observe an interesting change in proportion of macrostructure components in 2009.

Before 2009, the proportion of the \disc{} component was around 30\%, the \incomp{} component was stable in size, and the size of the \out{} component was increasing.
The real public adoption of Twitter started in 2009 where the total number of accounts went from 4.265 million in January 2009 to 67.487 million in January 2010.
Several events contributed to attract new users on Twitter during that period: the terrorist attacks in Mumbai was one of the first event followed on Twitter in November 2008, attracting the attention of other news media such as CNN; some influential celebrities started to use Twitter such as Oprah Winfrey, and, for the first time, some accounts reached one million followers.

We see in Figure~\ref{fig:evolution} (bottom) that the large adoption of Twitter in 2009 led to changes in the macrostructure of its social graph.
The proportion of the \disc{} component dropped to 10\% while the \lsc{} jumped to 70\%.
We have seen in Section~\ref{twitter2012} that the \disc{} component corresponds to abandoned accounts, so during such a large adoption phase, the proportion of abandoned accounts is much lower.
However, this proportion increased in 2010 and 2011 to reach a stable value, with the \disc{} component representing around 20\% of all accounts. 

We also observe in Figure~\ref{fig:evolution} (bottom) that the proportion of the \out{} component has been decreasing since 2009.
The reason is that a large fraction of celebrities joined Twitter in 2009 and 2010.
Some of these celebrities created an account to increase their visibility, but never intended to follow other accounts, thus they joined the \out{} component.
The fraction of such celebrities is decreasing compared to regular accounts, and also joining Twitter without following anyone in the \lsc{} component is a decreasing trend.
Indeed, the proportion of the \incomp{} component has been increasing since 2009, showing that it is an increasing trend to follow accounts in the \lsc{} component without tweeting and being followed.

It is worth noticing that the two most popular Twitter datasets~\cite{KwakEtAl2010,ChaEtAl2010} have been collected in 2009.
We have seen that the Twitter social graph macrostructure has significantly changed during the period 2009/2010, calling for a newer dataset such as the one we collected, which is more representative of the actual Twitter social graph.
We also note that the two datasets of 2009 do not contain accounts belonging to the \disc{} component, unlike our dataset, which is an issue for researchers focusing on malicious activities and abandoned accounts on Twitter. 

\begin{figure}[t]
\centering
\includegraphics[scale =0.44]{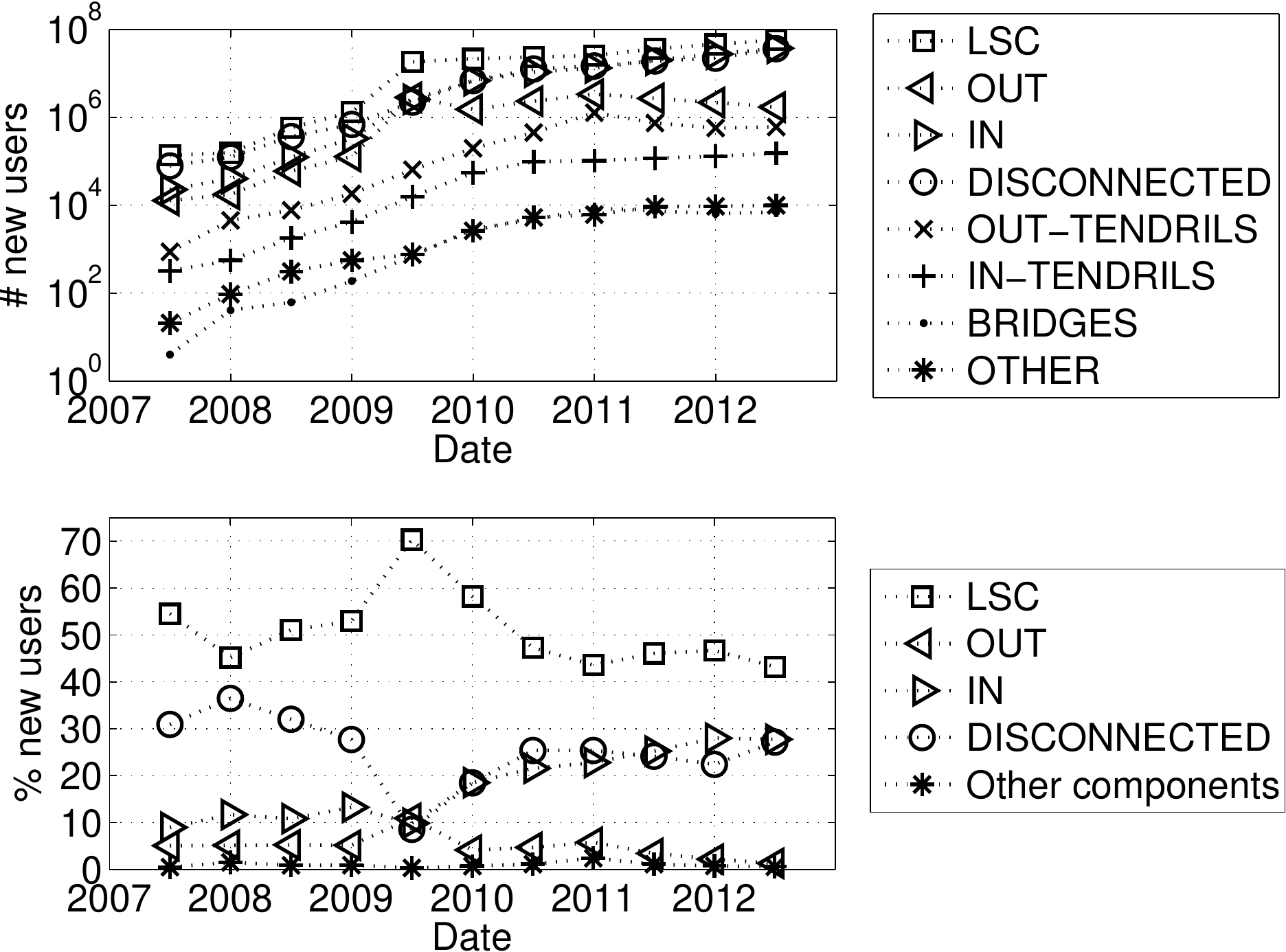} 
\caption{
\textbf{Distribution of new accounts per components with time.}
\textnormal{\sl (top) Number of new accounts per component.
(bottom) Fraction of the total number of new accounts per component.
}
}
\label{fig:new_users}
\end{figure}

\subsection{Distribution of New Accounts in Components }
\label{sec:repart-new-acco}

In this section, we evaluate to which component the new accounts created during each 6 months period belong to.
To find this distribution, we use the approximations of the Twitter social graph macrostructure described in Section~\ref{sec:evol-twitt-graph}.
Then, for each pair of contiguous approximations in time (e.g., July 2008 and January 2009), we remove all accounts already present in the oldest one to the newest one.
This way, we obtain the evolution with time of the distribution of the new accounts in components, see Figure~\ref{fig:new_users}. 

We observe in Figure~\ref{fig:new_users} (top) that the total number of new accounts increases with time for the \lsc{}, \incomp{}, and \disc{} components, but not for \out{}.
This decrease confirms our discussion in Section~\ref{sec:out-comp} on the \out{} component, explaining that new selfish accounts are decreasing in Twitter.

Figure~\ref{fig:new_users} (bottom) shows the fraction of the total number of new accounts per component.
We observe that new accounts join most the \lsc{} component, but this is a decreasing trend at the benefit of the \incomp{} and \disc{} components.
We explain this trend by two changes in the usage of Twitter initiated in 2009.
First, passive followers are taking an increasing role in Twitter; passive followers are accounts that follow other accounts, but that are not followed and never publish tweets, as described in Section~\ref{sec:in-comp}. 
This increasing role of passive followers shows that Twitter is more and more used as a regular information media in which people receive information, but do not produce any.
However, more than 40\% of new accounts are still joining the \lsc{} component, making Twitter the largest and most participative information media.
Second, as Twitter is very popular, it is attracting a large fraction of users that are just creating a Twitter account out of curiosity, but never effectively use it.
Most of these accounts end up in the \disc{} component. 

\section{Related work}
\label{sec:related-work}

Twitter has been widely studied for years.
A large fraction of the literature is on the identification of malicious behavior on Twitter \cite{thomas_suspended_2011, zhang_detecting_2011, ghosh_understanding_2012}, on the study of tweets propagation \cite{SadikovEtAl2009, YeEtAl2010}, and on privacy \cite{TwitterPrivacy1, TwitterPrivacy2}.
All these studies are not directly related to our work as they do not crawl the Twitter social graph and do not explore its properties.
\eject
Closer to our work, several studies focused on the Twitter social graph.
Some of them crawled partially the graph before 2009 \cite{JavaEtAl2007, KrishnamurthyEtAl2008, HubermanEtAl2008}, so before the wide adoption of Twitter. Two studies made a large crawl of the Twitter social graph.
Kwak \textit{et al.}\ used a technique close to a BFS and reverse BFS from a popular account and also collected accounts referring to trending topics.
This crawling methodology cannot capture some users that are not connected to the \lsc{} component, and that do not tweet about trending topic, thus a partial view of the Twitter social graph.
Cha \textit{et al.}\ \cite{ChaEtAl2010} used a crawl by account ID, that is close to what we did.
Both of these studies made their dataset publicly available and others built on it \cite{LeeEtAl2010, WuEtAl2011, ChaEtAl2012, sharma_inferring_2012}, but the datasets were collected in 2009 during the main change in the Twitter social graph we discussed in Section~\ref{sec:evol-twitt-graph}.

To the best of our knowledge, the dataset we present is the most up-to-date and the most complete description of the Twitter social graph.
Moreover, none of these studies explores the macrostructure of the Twitter social graph, a new way to represent directed social graphs.
Broder \textit{et al.} \cite{Broder2000} introduced first the notion of macrostructure for a directed graph in the context of the Web, but we significantly improved it, and we are the first ones to apply it to Twitter.
Unlike what Broder \textit{et al.} proposed, we present a methodology to compute the exhaustive macrostructure of any large directed social graph, along with the categorization of each account in the identified component, which is a significant methodological step.

\section{Conclusion}
\label{concl}

In this paper, we present the largest, most complete, and most up-to-date crawl of the Twitter social graph.
This graph contains 505 million accounts connected with 23 billion arcs.
In addition, we present a methodology to practically compute the macrostructure of any directed social graph and to exhaustively classify each account to one of the identified components.
We applied this methodology to the Twitter social graph and found that only 50.71\% of the accounts belong to the \lsc{} component, and that 21.60\% of the accounts (in the \disc{} component) have no path to the other accounts.

We show that the main components of the macrostructure of the Twitter social graph correspond to specific usages.
For instance, the \lsc{} component hold most of the regular Twitter activity, and the \incomp{} component holds passive followers.
Finally, we present a simple methodology to explore the evolution of the macrostructure of Twitter with time, we validate this methodology, and we show that the public datasets crawled in 2009 do not represent the current macrostructure of the Twitter social graph.

We believe that our collected dataset is a gold mine for any researcher working on social graphs and that the macrostructure analysis sheds a new light on the Twitter social graph that will be useful for both theoreticians and experimenters.

\section{Acknowledgements}
The authors thank Krishna P. Gummadi (MPI-SWS) for insights on the analysis of the dataset we collected and valuable feedback from the early stages of this work. We also thank him for sharing the list of influential Twitter users identified by Sharma \textit{et al.}~\cite{sharma_inferring_2012}.

\pagebreak[4]
\bibliographystyle{abbrv}

\end{document}